\documentclass[11pt]{article}
\usepackage{hyperref}
\usepackage{geometry}
\usepackage{graphicx}
\usepackage{amsmath,amssymb}
\usepackage{dsfont}
\usepackage{caption}
\usepackage{subcaption}
\usepackage{xcolor} %définir couleur à la main
\usepackage{cite} %gérer les citations proprement
\usepackage{feynmf}

\newcommand\pubnumber{}
\newcommand\pubdate{\today}

\def\Tub{Institut f\"{u}r Theoretische Physik, Eberhard Karls
  Universit\"{a}t T\"{u}bingen, Auf der Morgenstelle 14, D-72076
  T\"{u}bingen, Germany.}
\def\Pitt{Pittsburgh Particle physics, Astrophysics, and Cosmology Center,
Department of Physics and Astronomy, University of Pittsburgh,
3941 O'Hara St., Pittsburgh, PA 15260, USA}

\def\Title#1{\begin{center} {\Large #1 } \end{center}}
\def\Author#1{\begin{center}{ \sc #1} \end{center}}
\def\Address#1{\begin{center}{ \it #1} \end{center}}

\newcommand\pubblock{\rightline{\begin{tabular}{l} \pubnumber\\
         \pubdate  \end{tabular}}}
\newenvironment{Abstract}{\begin{quotation}  }{\end{quotation}}
\newenvironment{Presented}{\begin{quotation} \begin{center} 
             PRESENTED AT\end{center}\bigskip 
      \begin{center}\begin{large}}{\end{large}\end{center} \end{quotation}}
\def\Acknowledgements{\bigskip  \bigskip \begin{center} \begin{large}
             \bf ACKNOWLEDGEMENTS \end{large}\end{center}}
\def\beq{\begin{equation}}
\def\eeq#1{\label{#1}\end{equation}}
\def\eeqn{\end{equation}}

\def\beqa{\begin{eqnarray}}
\def\eeqa#1{\label{#1}\end{eqnarray}}
\def\eeqan{\end{eqnarray}}

\let\bar=\overbar

\def\Dslash{\not{\hbox{\kern-4pt $D$}}}
\def\dslash{\not{\hbox{\kern-2pt $\del$}}}

\def\msb{{\bar{\ssstyle M \kern -1pt S}}}

\begin{document}
\begin{titlepage}
\pubblock

\vfill
\Title{Searching for heavy neutrinos with WWH production}
\vfill
\Author{J. Baglio$^{\dagger}$, C. Weiland$^{\S,}$\footnote{Speaker}, }
\Address{$^\dagger$\Tub \\ $^\S$\Pitt}
\vfill
\begin{Abstract}
  Extensions of the Standard Model are required to give mass to the
  light neutrinos and explain neutrino oscillations. One of the
  simplest ideas is to introduce new heavy, gauge singlet fermions
  that play the role of right-handed neutrinos in a seesaw
  mechanism. They could have large Yukawa couplings to the Higgs
  boson, affecting the production of the Higgs bosons in association
  with a pair of W bosons at future lepton colliders. 
  Working in the inverse seesaw model and taking into account all
  possible experimental constraints, we find that sizable deviations,
  as large as 66\% are possible. This makes the $W^+_{}W^-_{}H$
  production cross-section a new, promising observable to constrain
  neutrino mass models. The effects are generic and expected to be
  present in other low-scale seesaw models.
\end{Abstract}
\vfill
\begin{Presented}
The International Workshop on Future Linear Colliders (LCWS2018),\\ Arlington, Texas, 22-26 October 2018. C18-10-22.
\end{Presented}
\vfill
\end{titlepage}
\def\thefootnote{\fnsymbol{footnote}}
\setcounter{footnote}{0}

\section{Introduction}

The observation of neutrino oscillations implies that at least two
neutrinos have a non-zero mass and that neutral lepton flavor is not
conserved~\cite{Esteban:2018azc}, thus requiring an extension of the
Standard Model (SM). Among the many ideas put forward to generate the
neutrino masses and mixing, one of the simplest is the addition of
right-handed neutrinos, which are fermionic gauge singlets. Including
all renormalizable terms allowed by the Standard Model gauge group
$\mathrm{SU}(3)_c\times \mathrm{SU}(2)_L \times U(1)_Y$ then naturally
leads to the type I
seesaw~\cite{Minkowski:1977sc,Ramond:1979py,GellMann:1980vs,Yanagida:1979as,Mohapatra:1979ia,Schechter:1980gr,Schechter:1981cv}. However
in this model, the size of 
light neutrino masses and the size of the new, heavy neutrinos
couplings to SM particles essentially depends on the same parameter,
suppressing them both and thus making it hard to experimentally probe 
this model. An appealing alternative is to consider low-scale seesaw
models, such as the inverse seesaw
model~\cite{Mohapatra:1986aw,Mohapatra:1986bd,Bernabeu:1987gr}, where
a nearly conserved symmetry is introduced. This symmetry, which has to
be lepton number~\cite{Kersten:2007vk,Moffat:2017feq}, allows large
couplings between heavy neutrinos and SM particles, which leads to a
rich phenomenology. In this talk, we discuss how the Higgs sector can
be used to probe these neutrino mass models, focusing on
$W^+_{}W^-_{}H$ production at lepton colliders.
While we present the results of a study in the inverse seesaw (ISS),
we expect our results to hold for other low-scale seesaw models. The
full details of this work can be found in the original
study~\cite{Baglio:2017fxf}.

\section{The inverse seesaw model: description and constraints}

In the realization of the inverse seesaw that we consider, each
generation is supplemented with a pair of right-handed gauge singlets,
$\nu_R^{}$ and $X$, with opposite lepton number.
The additional mass terms to the SM Lagrangian are given by
\begin{equation}
  \mathcal{L}_\mathrm{ISS} = - Y^{ij}_\nu \overline{L_{i}}
  \widetilde{\Phi} \nu_{Rj} - M_R^{ij} \overline{\nu_{Ri}^C} X_{Rj} -
  \frac{1}{2} \mu_{X}^{ij} \overline{X_{i}^C} X_{j} + \mathrm{
    h.c.}\,,
    \label{LagrangianISS}
\end{equation}
with $\Phi$ the SM Higgs field and $\widetilde \Phi=\imath \sigma_2
\Phi^*$, $i,j=1\dots 3$, $Y_\nu$ and $M_R$ complex matrices and
$\mu_{X}$ a complex symmetric matrix.
All terms are lepton number conserving, with the exception of the
naturally small $\mu_{X}$ to which the light neutrino masses are
proportional. Indeed for one generation and in the seesaw limit $\mu_X
\ll m_D, M_R$, the neutrino mass matrix admits as singular values
\begin{align}
 m_\nu &\simeq \frac{m_{D}^2}{m_{D}^2+M_{R}^2} \mu_X\,\label{mnu},\\
 m_{N_1,N_2} &\simeq \sqrt{M_{R}^2+m_{D}^2} \mp \frac{M_{R}^2 \mu_X}{2 (m_{D}^2+M_{R}^2)}\,,\label{mN}
\end{align}
where $m_D=Y_\nu \langle \Phi\rangle$. This corresponds to one light
neutrino and two heavy, nearly degenerate neutrinos with
opposite CP parities forming a pseudo-Dirac pair.
The smallness of $\mu_X$ allows to suppress the light neutrino mass
while keeping the mixing between active and sterile neutrinos (that is
proportional to $m_D M_R^{-1}$) large. As a consequence, it is
possible to have large Yukawa couplings even when the seesaw scale is
close to the electroweak scale.

Since a major motivation of these models is to explain neutrino
oscillations, we reproduce data from the global fit {NuFIT}
3.0~\cite{Esteban:2016qun} by using the $\mu_X$-parametrization with
next-order terms in the seesaw expansion that are relevant for large
active-sterile mixing~\cite{Baglio:2016bop}
\begin{align}
  \begin{split}
    \mu_X \simeq & \left(\mathbf{1}-\frac{1}{2} M_R^{*-1} m_D^\dagger
      m_D M_R^{T-1} \right)^{-1} M_R^T m_D^{-1} U_{\rm PMNS}^* m_\nu
    U_{\rm PMNS}^\dagger m_D^{T-1} M_R\, \\
    & \times \left(\mathbf{1}-\frac{1}{2} M_R^{-1} m_D^T m_D^*
      M_R^{\dagger-1}\right)^{-1}\,.
  \end{split}
    \label{muXparam}
\end{align}
Here, $m_\nu$ is the diagonal light neutrino mass matrix and $U_{\rm
  PMNS}$ is the unitary Pontecorvo-Maki-Nakagawa-Sakata
(PMNS)~\cite{Pontecorvo:1957cp,Maki:1962mu}. This parametrization uses
$Y_\nu$ and $M_R$ as input parameters. We will consider a scenario
where both of them are diagonal, suppressing the rates of
lepton-flavor-violating processes. Similarly constraints from the
electron electric dipole moment measurements are avoided by choosing
all mass matrices and couplings in the lepton sector to be real. As a
consequence, the strongest experimental constraints come from a global
fit~\cite{Fernandez-Martinez:2016lgt} to electroweak precision
observables, tests of CKM unitarity and tests of lepton
universality. Additionally, we require that the Yukawa couplings
$Y_\nu$ remain perturbative, namely
\begin{equation}
 \frac{|Y^{ij}_\nu|^2}{4 \pi} < 1.5\,.
\end{equation}

\section{Heavy neutrinos and $W^+_{}W^-_{}H$ production at lepton colliders}

With large neutrino Yukawa couplings and heavy neutrinos whose mass is
close to the electroweak scale, the Higgs sector appears as a prime
candidate to look for the imprint of the new particles present in the
inverse seesaw model. Off-diagonal couplings could give rise to
striking lepton-flavor-violating Higgs decays~\cite{Arganda:2014dta}
for example. If we restrict ourselves to seesaw scales above the Higgs
mass, heavy neutrinos can also induce large deviations in the Higgs
trilinear coupling~\cite{Baglio:2016ijw,Baglio:2016bop}. These can be as large as $10\%$
at an energy scale of 500 GeV, which would be within reach of future lepton colliders such
as the International Linear Collider at 1 TeV or CLIC 1.5 GeV, and even reach $30\%$
at an energy scale of 2.5 TeV, which would be testable at collider energies of 3 TeV and above. It is in
particular worth noting that these deviations are maximal for diagonal
neutrino Yukawa coupling and heavy neutrinos of masses around 10~TeV,
thus providing a new observable to test a regime very difficult to
probe otherwise.

\begin{figure}[!t]
  \centering
  \includegraphics[scale=0.53]{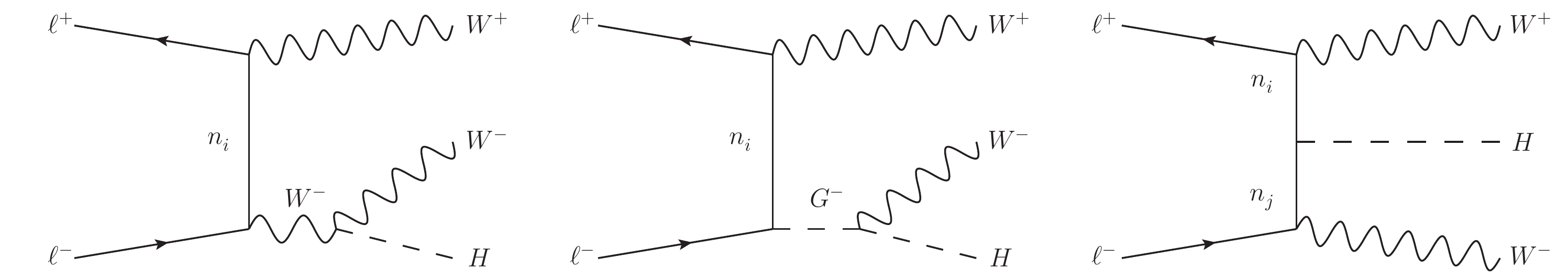}
  \caption[]{Feynman diagrams representing the ISS neutrino
    contributions to 
$\ell^+_{}\ell^-_{}\rightarrow W^+_{} W^-_{} H$
in the
    Feynman-'t~Hooft gauge. Mirror diagrams cam be obtained by
    flipping all the electric charges and the indices $i,j$ run from 1
    to 9.
  }
  \label{fig:feyndiags}
\end{figure}

Inspired by the observation that $t$--channel fermions coupled to a
Higgs boson can give sizable contributions to a cross-section, as is
the case of $b\bar{b}\rightarrow W^+_{} W^-_{} H$ at the LHC for
example~\cite{Baglio:2016ofi}, we turned our attention to the impact
of heavy neutrinos on the production of a Higgs boson in association
with a pair of $W$ bosons at a lepton collider, $e^+_{} e^-_{}
\rightarrow W^+_{} W^-_{} H$. Sensitivity studies in the SM reported
good detection prospects~\cite{Baillargeon:1993iw}. In the ISS, the additional contributions due to the $t$--channel exchange
of massive neutrinos, with respect to SM contributions, are
given by the diagrams of fig.~\ref{fig:feyndiags}.
Details of the calculation as well as values for the SM and neutrino
inputs can be found in our original study~\cite{Baglio:2017fxf}.

\begin{figure}[!t]
 \centering
 \includegraphics[width=0.44\textwidth]{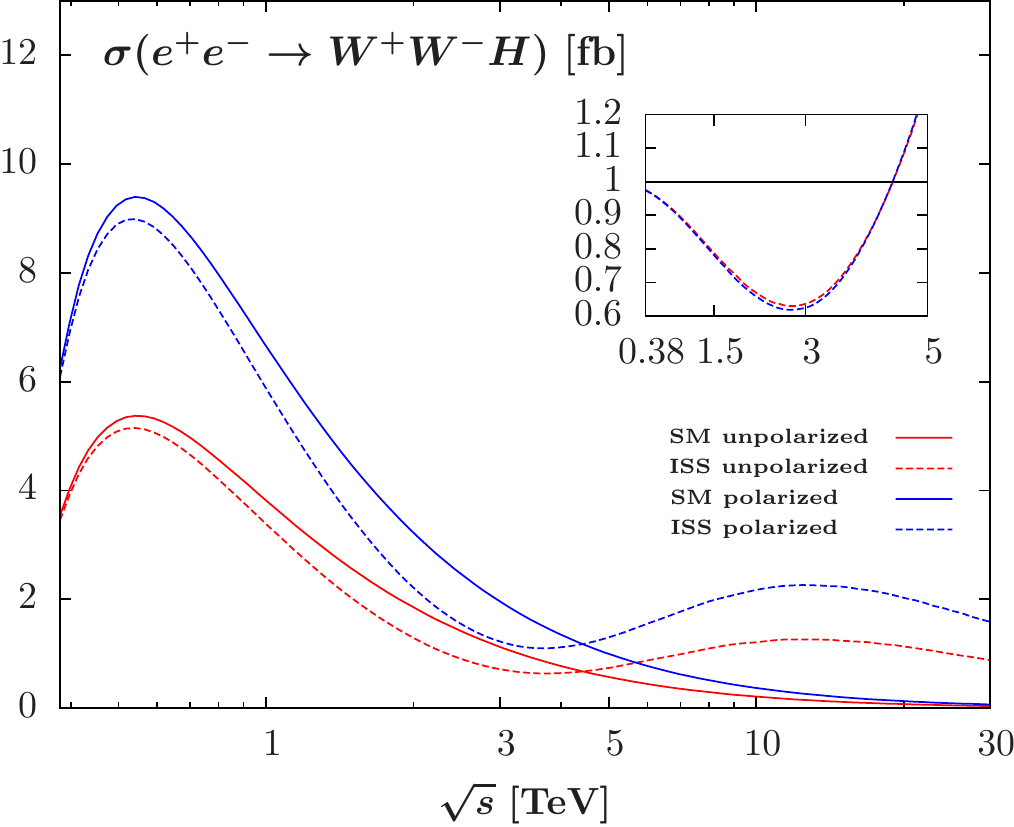}
\hspace{5mm}
 \includegraphics[width=0.50\textwidth]{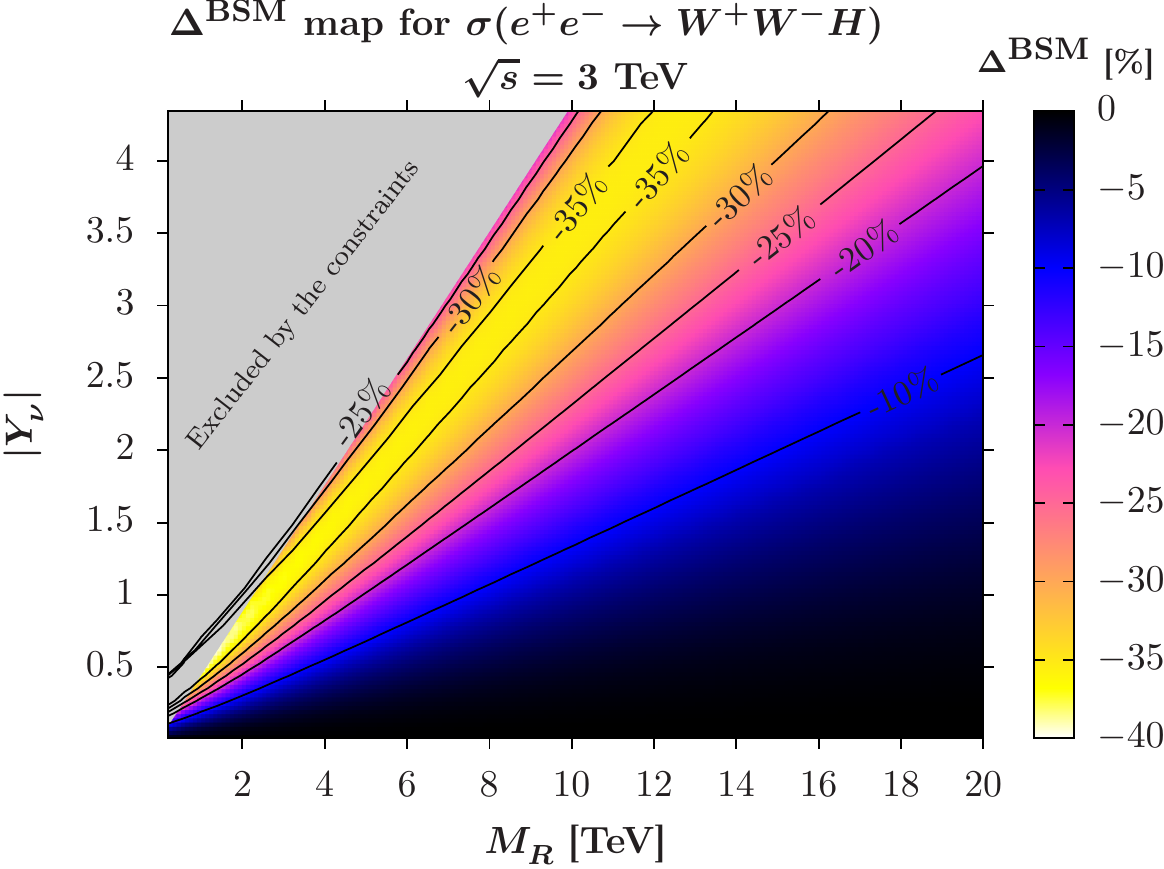}
\caption[]{Left: Leading order total cross-section as a function of
  the collider energy $\sqrt{s}$ (taken
    from~\cite{deBlas:2018mhx}). The SM predictions are given by the
  solid curves while the dashed curves 
  stand for the ISS predictions using the benchmark scenario  defined in the text. The red (blue)
  curves are for an unpolarized ($-80\%$ polarized electron beam)
  cross-sections. The insert displays the ratio of the ISS prediction
  with respect to the SM cross-section up to 5 TeV.
  Right: Contour map of the neutrino corrections $\Delta_{}^{\rm BSM}$
  at the 3 TeV CLIC, using a $-80\%$ polarized electron beam, as a
  function of the seesaw scale $M_R^{}$ and $|Y_\nu^{}|$
    (taken from~\cite{Baglio:2017fxf}).}
\label{fig:xs}
\end{figure}

In the fig.~\ref{fig:xs} (left),
we present our numerical results for a benchmark scenario with $Y_\nu
= \mathds{1}$ and a hierarchical heavy neutrino spectrum where the
masses of the pseudo-Dirac pairs are respectively 2.4~TeV, 3.6~TeV (this is the
pseudo-Dirac pair that couples to the electron) and 8.6~TeV. We present results up
  to center-of-mass energies of 30 TeV. Since 
the process is very sensitive to the chirality of the incoming
electron and positron, favoring left-handed electrons and right-handed
positrons, we compare the polarized and unpolarized
cross-sections, choosing the Compact Linear Collider (CLIC)
baseline~\cite{CLIC:2016zwp}. It has an unpolarized positron beam,
$P_{e^+_{}}^{} = 0$, and a polarized electron beam with
$P_{e^-_{}}^{}=-80\%$. First, we observe that the polarized
cross-section is nearly twice the unpolarized one, demonstrating the
dependence on the electron chirality mentioned above.
Second, we can see below 4~TeV that the presence of additional, heavy
neutrinos reduces the cross-section. This is due to a destructive
interference between the $s$--channel and $t$--channel diagrams which
is already present in the SM and is exacerbated in the ISS. In this
regime, the deviation is maximal close to 3~TeV and reaches
$-38\%$. Third, if we keep increasing the center-of-mass energy, the
intermediate heavy neutrino gets closer and closer to being on-shell,
leading the heavy neutrino diagrams to dominate the amplitude and to a
subsequent increase and large enhancement of the cross-section. This
regime could typically be probed at very-high-energy lepton
colliders based on different accelerator technologies, using muon
beams like LEMMA~\cite{Shiltsev:2018qbd} or high gradient acceleration
concepts such as ALIC~\cite{ALEGRO:2019alc}.

Fig.~\ref{fig:xs} (right) shows a contour map of the deviation of the
ISS cross-section with respect to that of the SM,
$\Delta^{\rm BSM}_{} = (\sigma^{\rm ISS}_{}-\sigma^{\rm
  SM}_{})/\sigma^{\rm SM}_{}$, as a function of the seesaw scale $M_R$
and of the neutrino Yukawa  coupling $Y_\nu$, at the 3 TeV CLIC with a
-80\% polarized electron beam. We are working with a hierarchical
spectrum for the heavy neutrinos that preserves the ratios used  for
the benchmark of fig.~\ref{fig:xs} (left), $M_{R_1^{}}^{} = 1.51
M_R^{}$, $M_{R_2^{}}^{} = 3.59 M_R^{}$ and $M_{R_3^{}}^{} =
M_R^{}$. The gray area excluded by the constraints mostly originates
from the global fit~\cite{Fernandez-Martinez:2016lgt}. The largest
deviation in the ISS reaches $-38\%$ for $|Y_\nu^{}|\sim 1$ and a
seesaw scale of a few TeV. These results can be approximated within
$1\%$ for $M_R^{} > 3$~TeV by using the formula 
\begin{align}
  \mathcal{A}^{\rm ISS}_{\rm approx} =\, 
  & \frac{(1~\text{TeV})_{}^2}{M_R^2} {\rm   Tr} (Y_\nu^{}
    Y_\nu^\dagger)\, \left(17.07 -
    \frac{19.79~\text{TeV}_{}^2}{M_R^2}\right),\nonumber\\
 \Delta^{\rm BSM}_{\rm approx} =\, 
  & (\mathcal{A}^{\rm ISS}_{\rm approx})_{}^2 -11.94\,
    \mathcal{A}^{\rm ISS}_{\rm approx}.
 \label{eq:approx}
\end{align}
We can see here that the process $e^+_{} e^-_{} \rightarrow W^+_{}
W^-_{} H$ exhibits sizable deviations of at least $-20\%$ for a large
fraction of the parameter space. It is worth comparing this result to
the one we obtained for the trilinear Higgs coupling
in~\cite{Baglio:2016bop}. When doing so, it is possible to see that
sizable deviations can be obtained for $W^+_{}W^-_{}H$ production in
a much larger part of the parameter space than for the trilinear Higgs
coupling.

\begin{figure*}[t]
  \centering
  \includegraphics[scale=0.77]{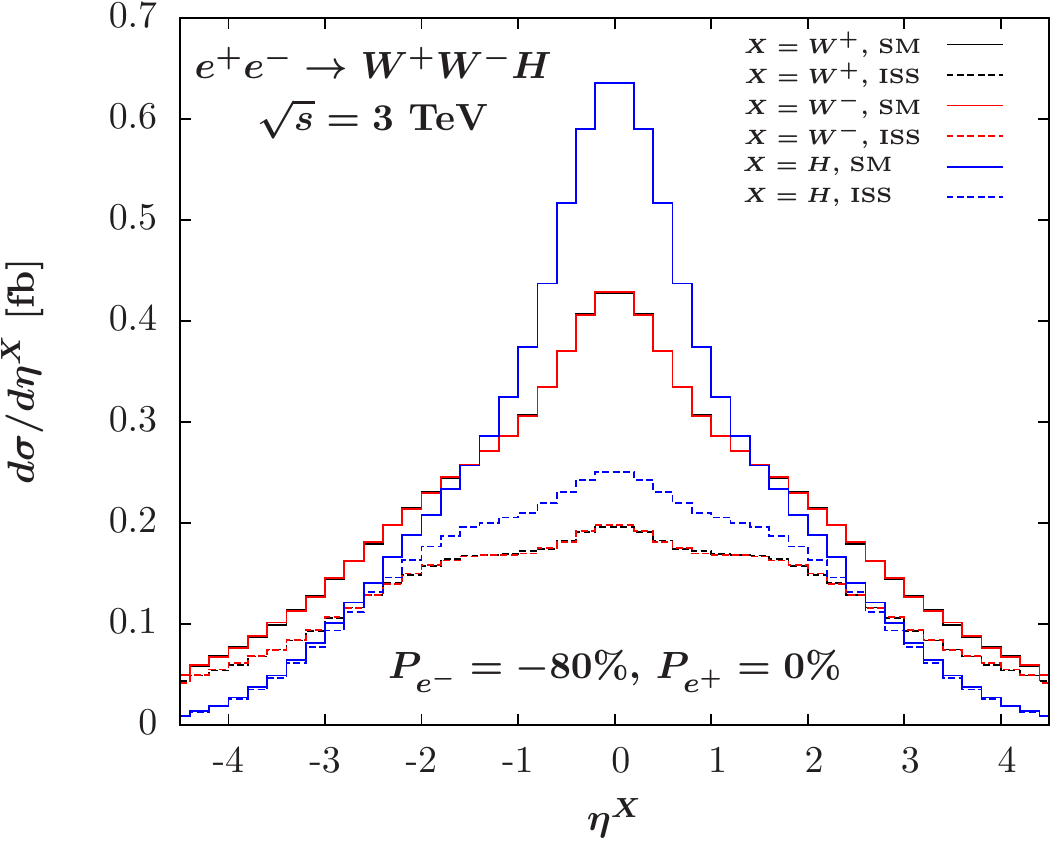}
  \hspace{3mm}
  \includegraphics[scale=0.77]{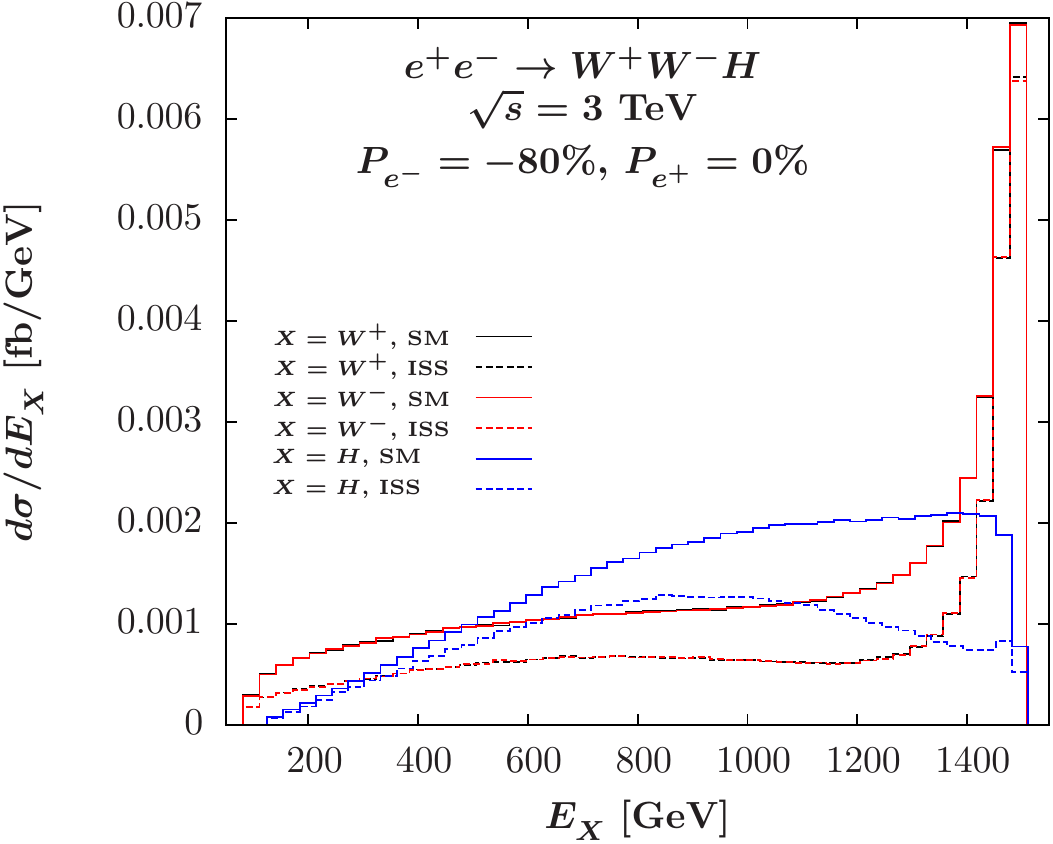}
  \caption[]{Pseudo-rapidity (left) and energy (right) distributions
    of the $W^+_{}$ (black), $W^-_{}$ (red) and Higgs (blue) bosons at
    $\sqrt{s}=3$ TeV, with a $-80\%$ polarized electron beam. The
    solid curves are for the SM predictions while the dashed curves
    are for the ISS  predictions using the same benchmark scenario as
    fig.~\ref{fig:xs} (left).}
  \label{fig:xsdists}
\end{figure*}

Fig.~\ref{fig:xsdists} presents the kinematic distribution in
pseudo-rapidity and energy of the $W$ and Higgs bosons at a
  center-of-mass energy of 3 TeV.
We can clearly see that the shape of the SM and ISS distributions are
easily distinguishable, with a noticeable difference in the central
region and for boosted Higgs bosons. As a consequence, the deviation
$\Delta^{\rm BSM}_{}$ can be enhanced with a simple choice of
cuts. It
was found that choosing $|\eta_{H/W^\pm_{}}^{}| < 1$ and $E_H^{} >
1$~TeV pushes the corrections down to $-66\%$ without decreasing the
cross-section by more than an order of magnitude. Indeed the ISS
cross-section after cuts was found to be $0.14$~fb, which has to be
compared to $1.23$~fb before cuts.

\section{Conclusion}

Low-scale seesaw models provide appealing extensions of the SM that
can generate neutrino masses and mixing and which are potentially
testable at colliders and in low-energy experiments. Having new
particles with a mass close to the Higgs mass and large Yukawa
coupling, the can naturally leave a wide imprint on the Higgs
sector. We have presented here the results of our
study~\cite{Baglio:2017fxf} where we considered the effect of heavy
neutrinos in the inverse seesaw on the production cross-section of
$\ell^+_{} \ell^-_{}\to W^+_{} W^-_{} H$. We found that at tree-level
corrections as large as $-38\%$ to the total cross-section can be
obtained at CLIC, which can be enhanced to $-66\%$ after applying very
basic cuts. Besides the deviations are found to be sizable in a
significant fraction of the parameter and we expect our results to
hold for other low-scale seesaw models. This makes this process a new
probe of neutrino mass model, allowing to test regimes with diagonal
and real Yukawa couplings which are difficult to access
otherwise. This is highly complementary to other existing probes such
as lepton-flavor-violating processes in the $\mathcal{O}(10)$~TeV
range.

\Acknowledgements

J.~B. acknowledges the support from the Institutional Strategy of the
University of T\"ubingen (DFG, ZUK 63), the DFG Grant JA 1954/1, and
the Carl-Zeiss foundation. C.W. received financial support from the
European Research Council under the European Union’s   Seventh
Framework Programme (FP/2007-2013)/ERC Grant NuMass Agreement
No. 617143. C.W. is also supported in part by the U.S.~Department of
Energy under contract DE-FG02-95ER40896 and in part by the PITT PACC.

\bibliographystyle{JHEP}
\bibliography{LCWS_Baglio_Weiland}

\providecommand{\href}[2]{#2}\begingroup\raggedright\begin{thebibliography}{10}

\bibitem{Esteban:2018azc}
I.~Esteban, M.~C. Gonzalez-Garcia, A.~Hernandez-Cabezudo, M.~Maltoni and
  T.~Schwetz, \emph{{Global analysis of three-flavour neutrino oscillations:
  synergies and tensions in the determination of $\theta_{23}, \delta_{CP}$,
  and the mass ordering}},
  \href{http://dx.doi.org/10.1007/JHEP01(2019)106}{\emph{JHEP} {\bf 01} (2019)
  106}, [\href{http://arxiv.org/abs/1811.05487}{{\tt 1811.05487}}].

\bibitem{Minkowski:1977sc}
P.~Minkowski, \emph{{$\mu \to e\gamma$ at a Rate of One Out of $10^{9}$ Muon
  Decays?}}, \href{http://dx.doi.org/10.1016/0370-2693(77)90435-X}{\emph{Phys.
  Lett.} {\bf B67} (1977) 421--428}.

\bibitem{Ramond:1979py}
P.~Ramond, \emph{{The Family Group in Grand Unified Theories}},  in
  \emph{{International Symposium on Fundamentals of Quantum Theory and Quantum
  Field Theory Palm Coast, Florida, February 25-March 2, 1979}}, pp.~265--280,
  1979.
\newblock \href{http://arxiv.org/abs/hep-ph/9809459}{{\tt hep-ph/9809459}}.

\bibitem{GellMann:1980vs}
M.~Gell-Mann, P.~Ramond and R.~Slansky, \emph{{Complex Spinors and Unified
  Theories}}, {\emph{Conf. Proc.} {\bf C790927} (1979) 315--321},
  [\href{http://arxiv.org/abs/1306.4669}{{\tt 1306.4669}}].

\bibitem{Yanagida:1979as}
T.~Yanagida, \emph{{Horizontal symmetry and masses of neutrinos}}, {\emph{Conf.
  Proc.} {\bf C7902131} (1979) 95--99}.

\bibitem{Mohapatra:1979ia}
R.~N. Mohapatra and G.~Senjanovic, \emph{{Neutrino Mass and Spontaneous Parity
  Violation}}, \href{http://dx.doi.org/10.1103/PhysRevLett.44.912}{\emph{Phys.
  Rev. Lett.} {\bf 44} (1980) 912}.

\bibitem{Schechter:1980gr}
J.~Schechter and J.~W.~F. Valle, \emph{{Neutrino Masses in $SU(2) \times U(1)$
  Theories}}, \href{http://dx.doi.org/10.1103/PhysRevD.22.2227}{\emph{Phys.
  Rev.} {\bf D22} (1980) 2227}.

\bibitem{Schechter:1981cv}
J.~Schechter and J.~W.~F. Valle, \emph{{Neutrino Decay and Spontaneous
  Violation of Lepton Number}},
  \href{http://dx.doi.org/10.1103/PhysRevD.25.774}{\emph{Phys. Rev.} {\bf D25}
  (1982) 774}.

\bibitem{Mohapatra:1986aw}
R.~N. Mohapatra, \emph{{Mechanism for Understanding Small Neutrino Mass in
  Superstring Theories}},
  \href{http://dx.doi.org/10.1103/PhysRevLett.56.561}{\emph{Phys. Rev. Lett.}
  {\bf 56} (1986) 561--563}.

\bibitem{Mohapatra:1986bd}
R.~N. Mohapatra and J.~W.~F. Valle, \emph{{Neutrino Mass and Baryon Number
  Nonconservation in Superstring Models}},
  \href{http://dx.doi.org/10.1103/PhysRevD.34.1642}{\emph{Phys. Rev.} {\bf D34}
  (1986) 1642}.

\bibitem{Bernabeu:1987gr}
J.~{Bernab\'eu}, A.~Santamaria, J.~Vidal, A.~Mendez and J.~W.~F. Valle,
  \emph{{Lepton Flavor Nonconservation at High-Energies in a Superstring
  Inspired Standard Model}},
  \href{http://dx.doi.org/10.1016/0370-2693(87)91100-2}{\emph{Phys. Lett.} {\bf
  B187} (1987) 303}.

\bibitem{Kersten:2007vk}
J.~Kersten and A.~{\relax Yu}. Smirnov, \emph{{Right-Handed Neutrinos at CERN
  LHC and the Mechanism of Neutrino Mass Generation}},
  \href{http://dx.doi.org/10.1103/PhysRevD.76.073005}{\emph{Phys. Rev.} {\bf
  D76} (2007) 073005}, [\href{http://arxiv.org/abs/0705.3221}{{\tt
  0705.3221}}].

\bibitem{Moffat:2017feq}
K.~Moffat, S.~Pascoli and C.~Weiland, \emph{{Equivalence between massless
  neutrinos and lepton number conservation in fermionic singlet extensions of
  the Standard Model}},  \href{http://arxiv.org/abs/1712.07611}{{\tt
  1712.07611}}.

\bibitem{Baglio:2017fxf}
J.~Baglio, S.~Pascoli and C.~Weiland, \emph{{$W^+ W^- H$ production at lepton
  colliders: a new hope for heavy neutral leptons}},
  \href{http://dx.doi.org/10.1140/epjc/s10052-018-6279-x}{\emph{Eur. Phys. J.}
  {\bf C78} (2018) 795}, [\href{http://arxiv.org/abs/1712.07621}{{\tt
  1712.07621}}].

\bibitem{Esteban:2016qun}
I.~Esteban, M.~C. Gonzalez-Garcia, M.~Maltoni, I.~Martinez-Soler and
  T.~Schwetz, \emph{{Updated fit to three neutrino mixing: exploring the
  accelerator-reactor complementarity}},
  \href{http://dx.doi.org/10.1007/JHEP01(2017)087}{\emph{JHEP} {\bf 01} (2017)
  087}, [\href{http://arxiv.org/abs/1611.01514}{{\tt 1611.01514}}].

\bibitem{Baglio:2016bop}
J.~Baglio and C.~Weiland, \emph{{The triple Higgs coupling: A new probe of
  low-scale seesaw models}},
  \href{http://dx.doi.org/10.1007/JHEP04(2017)038}{\emph{JHEP} {\bf 04} (2017)
  038}, [\href{http://arxiv.org/abs/1612.06403}{{\tt 1612.06403}}].

\bibitem{Pontecorvo:1957cp}
B.~Pontecorvo, \emph{{Mesonium and anti-mesonium}}, {\emph{Sov. Phys. JETP}
  {\bf 6} (1957) 429}.

\bibitem{Maki:1962mu}
Z.~Maki, M.~Nakagawa and S.~Sakata, \emph{{Remarks on the unified model of
  elementary particles}},
  \href{http://dx.doi.org/10.1143/PTP.28.870}{\emph{Prog. Theor. Phys.} {\bf
  28} (1962) 870--880}.

\bibitem{Fernandez-Martinez:2016lgt}
E.~Fernandez-Martinez, J.~Hernandez-Garcia and J.~Lopez-Pavon, \emph{{Global
  constraints on heavy neutrino mixing}},
  \href{http://dx.doi.org/10.1007/JHEP08(2016)033}{\emph{JHEP} {\bf 08} (2016)
  033}, [\href{http://arxiv.org/abs/1605.08774}{{\tt 1605.08774}}].

\bibitem{Arganda:2014dta}
E.~Arganda, M.~J. Herrero, X.~Marcano and C.~Weiland, \emph{{Imprints of
  massive inverse seesaw model neutrinos in lepton flavor violating Higgs boson
  decays}}, \href{http://dx.doi.org/10.1103/PhysRevD.91.015001}{\emph{Phys.
  Rev.} {\bf D91} (2015) 015001}, [\href{http://arxiv.org/abs/1405.4300}{{\tt
  1405.4300}}].

\bibitem{Baglio:2016ijw}
J.~Baglio and C.~Weiland, \emph{{Heavy neutrino impact on the triple Higgs
  coupling}}, \href{http://dx.doi.org/10.1103/PhysRevD.94.013002}{\emph{Phys.
  Rev.} {\bf D94} (2016) 013002}, [\href{http://arxiv.org/abs/1603.00879}{{\tt
  1603.00879}}].

\bibitem{Baglio:2016ofi}
J.~Baglio, \emph{{Gluon fusion and $b\bar{b}$ corrections to $H W^+ W^- / H Z
  Z$ production in the POWHEG-BOX}},
  \href{http://dx.doi.org/10.1016/j.physletb.2016.10.066}{\emph{Phys. Lett.}
  {\bf B764} (2017) 54--59}, [\href{http://arxiv.org/abs/1609.05907}{{\tt
  1609.05907}}].

\bibitem{Baillargeon:1993iw}
M.~Baillargeon, F.~Boudjema, F.~Cuypers, E.~Gabrielli and B.~Mele, \emph{{Higgs
  production in association with a vector boson pair at future e+ e-
  colliders}},
  \href{http://dx.doi.org/10.1016/0550-3213(94)90298-4}{\emph{Nucl. Phys.} {\bf
  B424} (1994) 343--373}, [\href{http://arxiv.org/abs/hep-ph/9307225}{{\tt
  hep-ph/9307225}}].

\bibitem{deBlas:2018mhx}
J.~de~Blas et~al., \emph{{The CLIC Potential for New Physics}},
  \href{http://arxiv.org/abs/1812.02093}{{\tt 1812.02093}}.

\bibitem{CLIC:2016zwp}
{\scshape CLICdp, CLIC} collaboration, M.~J. Boland et~al., \emph{{Updated
  baseline for a staged Compact Linear Collider}},
  \href{http://arxiv.org/abs/1608.07537}{{\tt 1608.07537}}.

\bibitem{Shiltsev:2018qbd}
V.~Shiltsev and D.~Neuffer, \emph{{On the Feasibility of a Pulsed 14 TeV C.M.E.
  Muon Collider in the LHC Tunnel}},  in \emph{{Proceedings, 9th International
  Particle Accelerator Conference (IPAC 2018): Vancouver, BC Canada}},
  p.~MOPMF072, 2018.
\newblock \href{http://dx.doi.org/10.18429/JACoW-IPAC2018-MOPMF072}{DOI}.

\bibitem{ALEGRO:2019alc}
{\scshape ALEGRO} collaboration, \emph{{Towards an Advanced Linear
  International Collider}},  \href{http://arxiv.org/abs/1901.10370}{{\tt
  1901.10370}}.

\end{thebibliography}\endgroup

\end{document}